\documentclass{jpsj-suppl}
\usepackage{txfonts, bm, amsmath} 

\usepackage[dvipdfmx]{color}

\newcommand{\Psfig}[2]{\includegraphics[width=#1]{#2}}

\def\mev{\text{ MeV}}
\def\gev{\text{ GeV}}
\def\fm{\text{ fm}}

\title{Dynamically Generated $\Xi (1690)$}

\author{Takayasu \textsc{Sekihara}$^{1}$}

\inst{$^{1}$Research Center for Nuclear Physics (RCNP), Osaka
  University, Ibaraki, Osaka, 567-0047, Japan}

\email{sekihara@rcnp.osaka-u.ac.jp}

\recdate{December 25, 2015}

\abst{We show that the $\Xi (1690)$ resonance can be dynamically
  generated in the $s$-wave $\bar{K} \Sigma$-$\bar{K} \Lambda$-$\pi
  \Xi$-$\eta \Xi$ coupled-channels chiral unitary approach.  In our
  model, the $\Xi (1690)$ resonance appears near the $\bar{K} \Sigma$
  threshold as a $\bar{K} \Sigma$ molecular state and the experimental
  data are reproduced well.  We discuss properties of the dynamically
  generated $\Xi (1690)$.}

\kword{$\Xi (1690)$, chiral unitary approach, hadronic molecules}

\begin{document}
\maketitle

\section{Introduction}

At present, spectroscopy of multi-strangeness baryons is not well
understood compared to that with the strangeness $S = 0$ and $-
1$~\cite{Agashe:2014kda}.  Nevertheless, we expect that
multi-strangeness baryons should contain interesting physics as rich
as in the baryons with $S = 0$ and $- 1$.  For instance, several
multi-strangeness baryons would contradict the classification with the
$q q q$ configuration by traditional quark models, and hence they
would be candidates of the exotic hadrons, like the famous $\Lambda
(1405)$ resonance in $S = -1$.  Therefore, investigating the baryon
spectroscopy in the multi-strangeness sector is important both in the
experimental and theoretical sides.

Here we focus on the $\Xi (1690)$ resonance.  The $\Xi (1690)$
resonance was experimentally discovered in the $K^{-} p \to ( \bar{K}
\Sigma ) K \pi$ reaction at $4.2 \gev / c$~\cite{Dionisi:1978tg},
which was followed by experimental studies in, {\it e.g.},
Refs.~\cite{Biagi:1981cu, Biagi:1986zj, Adamovich:1997ud, Abe:2001mb,
  Link:2005ut, Aubert:2008ty, Ablikim:2015apm}.  The spin/parity of
$\Xi (1690)$ has not been fixed yet, but experimental data prefer
$J^{P} = 1/2^{-}$~\cite{Dionisi:1978tg, Aubert:2008ty}.  Then, one of
the most interesting properties of $\Xi (1690)$ is its decay pattern.
Namely, $\Xi (1690)$ has small decay width $\lesssim 10
\mev$~\cite{Adamovich:1997ud} with tiny branching ratio to the $\pi
\Xi$ channel, $\Gamma ( \pi \Xi ) / \Gamma ( \bar{K} \Sigma ) <
0.09$~\cite{Agashe:2014kda}.  These experimental implications
contradict a naive quark model, in which a $q s s$ state with $J^{P} =
1/2^{-}$ should inevitably decay to $\pi \Xi$ to some extent.  These
properties of the decay pattern may imply that $\Xi (1690)$ might be
an exotic hadron rather than a usual $q q q$ baryon.

In this Article we show that the $\Xi (1690)$ resonance can be
dynamically generated as an $s$-wave $\bar{K} \Sigma$ molecular state
in the chiral unitary approach.  Furthermore, we solve the difficulty
on the decay pattern of $\Xi (1690)$ with $J^{P} = 1/2^{-}$ by using
the chiral unitary approach.  We note that the multi-strangeness
baryons were studied in the chiral unitary approach in
Refs.~\cite{Ramos:2002xh, GarciaRecio:2003ks, Gamermann:2011mq}.  In
this study we concentrate on the phenomena around the $\bar{K} \Sigma$
threshold and $\Xi (1690)$~\cite{Sekihara:2015qqa}.

\section{Dynamically generated $\Xi (1690)$}

In the chiral unitary approach, we calculate the meson--baryon
scattering amplitude $T_{j k}$ with the channel indices $j$ and $k$ by
the Lippmann--Schwinger equation in an algebraic form
\begin{equation}
  T_{j k} ( w ) = V_{j k} ( w ) 
  + \sum _{l} V_{j l} ( w ) G_{l} ( w ) T_{l k} ( w ) ,
\end{equation}
where $w$ is the center-of-mass energy, $V_{j k}$ is the interaction
kernel taken from the chiral perturbation theory, and $G_{j}$ is the
two-body loop function.  In this study, we take the Weinberg--Tomozawa
term for the interaction kernel $V_{j k}$.  For the loop function
$G_{j}$, we employ a covariant expression and evaluate it with the
dimensional regularization, which brings us a subtraction constant as
a model parameter in each channel.  We note that only the subtraction
constants are model parameters in the present study.  The details of
the formulation can be found in Ref.~\cite{Sekihara:2015qqa}.

\begin{figure}[t]
  \centering
  \Psfig{7.6cm}{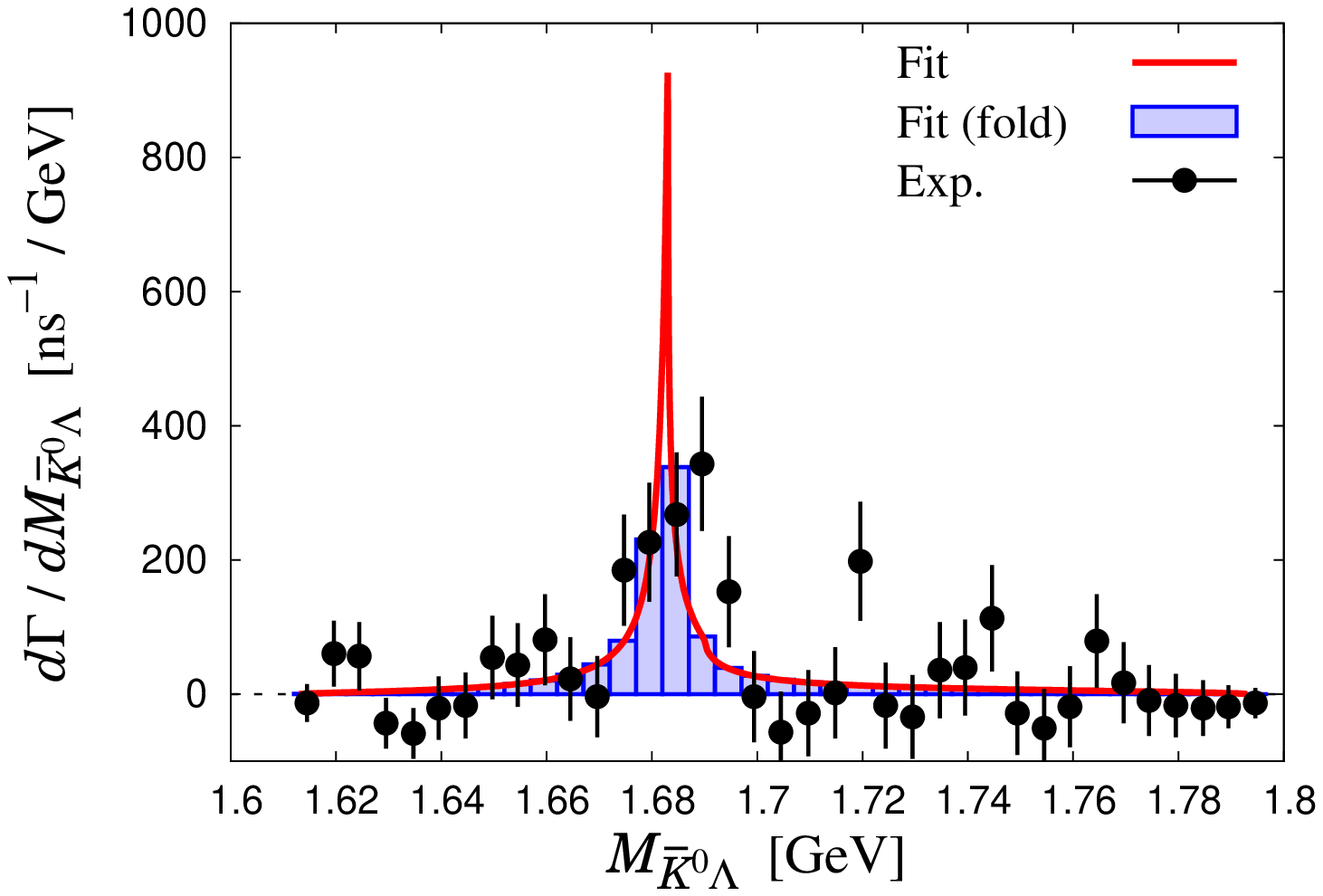} ~ \Psfig{7.6cm}{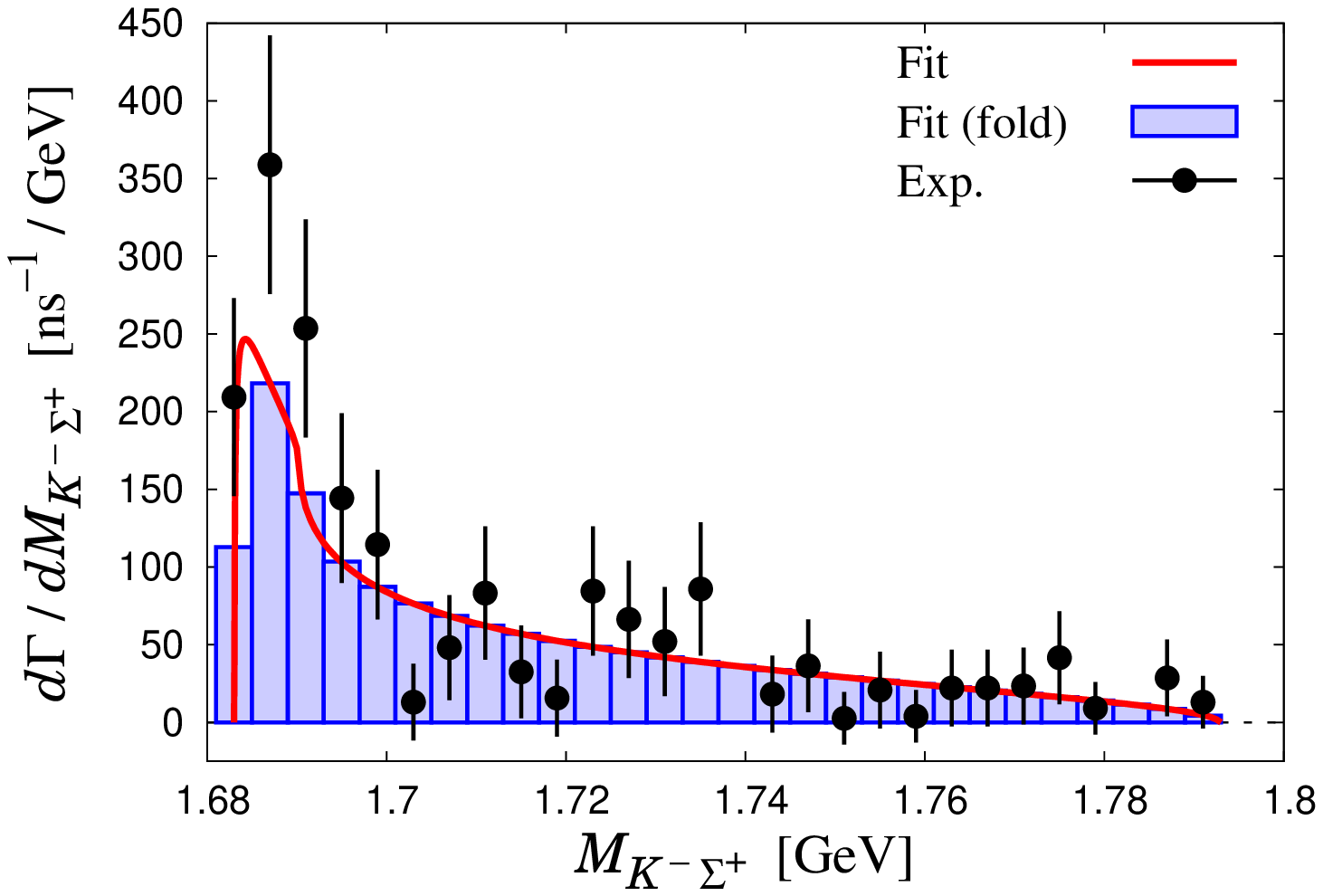}
  \caption{Mass spectra of $\bar{K}^{0} \Lambda$ (left) and $K^{-}
    \Sigma ^{+}$ (right)~\cite{Sekihara:2015qqa}.  The red lines and
    blue histograms are the theoretical mass spectra in the present
    model and its folded results with the size of experimental bins.
    The experimental data are taken from Ref.~\cite{Abe:2001mb}. }
  \label{fig:1}
\end{figure}

The parameters are fixed so as to reproduce simultaneously the
$\bar{K}^{0} \Lambda$ and $K^{-} \Sigma ^{+}$ mass spectra obtained in
the decay $\Lambda _{c}^{+} \to \Xi (1690)^{0} K^{+} \to ( \bar{K}^{0}
\Lambda ) K^{+}$ and $( K^{-} \Sigma ^{+}) K^{+}$ in
Ref.~\cite{Abe:2001mb}.  From the best fit, we obtain the mass spectra
in Fig.~\ref{fig:1}.  From the figure, we can reproduce the
experimental peak structure around the $\bar{K} \Sigma$ threshold
($\approx 1690 \mev$).  In addition, the ratio $R \equiv \mathcal{B} [
\Lambda _{c}^{+} \to \Xi (1690)^{0} K^{+} \to ( K^{-} \Sigma ^{+} )
K^{+} ] / \mathcal{B} [\Lambda _{c} \to \Xi (1690)^{0} K^{+} \to (
\bar{K}^{0} \Lambda ) K^{+} ]$ is evaluated as $1.06$ in our model,
which is within $2 \sigma$ of the experimental value $0.62 \pm
0.33$~\cite{Agashe:2014kda}.

\begin{table}[b]
  \caption{Properties of $\Xi (1690)^{0}$ in the present 
    model~\cite{Sekihara:2015qqa}.  The 
    pole position is $w_{\rm pole} = 1684.3 - 0.5 i \mev$.}
  \label{tab:1}
  \begin{tabular}{lcclc}
    \hline
    $g_{K^{-} \Sigma ^{+}}$ &
    $1.02 + 0.60 i$ 
    & &
    $X_{K^{-} \Sigma ^{+}}$ &
    $0.83 - 0.31 i$
    \\
    $g_{\bar{K}^{0} \Sigma ^{0}}$ &
    $-0.76 - 0.41 i \phantom{-}$
    & &
    $X_{\bar{K}^{0} \Sigma ^{0}}$ &
    $0.12 + 0.17 i$
    \\
    $g_{\bar{K}^{0} \Lambda}$ &
    $0.38 + 0.20 i$ 
    & &
    $X_{\bar{K}^{0} \Lambda}$ &
    $-0.02 + 0.00 i \phantom{-}$
    \\
    $g_{\pi ^{+} \Xi ^{-}}$ &
    $0.06 - 0.05 i$
    & &
    $X_{\pi ^{+} \Xi ^{-}}$ &
    $0.00 + 0.00 i$
    \\
    $g_{\pi ^{0} \Xi ^{0}}$ &
    $-0.09 + 0.05 i \phantom{-}$ 
    & &
    $X_{\pi ^{0} \Xi ^{0}}$ &
    $0.00 + 0.00 i$ 
    \\
    $g_{\eta \Xi ^{0}}$ &
    $-0.66 - 0.48 i \phantom{-}$
    & &
    $X_{\eta \Xi ^{0}}$ &
    $0.01 + 0.02 i$
    \\
    & & &
    $Z$ 
    & $0.06 + 0.11 i$ 
    \\
    \hline
  \end{tabular}
\end{table}

Next, in the scattering amplitude a resonance appears as a pole in the
following expression:
\begin{equation}
  T_{j k} ( w ) = \frac{g_{j} g_{k}}{w - w_{\rm pole}} 
  + (\text{regular at } w_{\rm pole}) ,
  \label{eq:amp}
\end{equation}
where $g_{j}$ is the coupling constant of the resonance to the
scattering state in $j$ channel and $w_{\rm pole}$ is the resonance
pole position.  In our model we find the $\Xi (1690)^{0}$ resonance
pole at $w_{\rm pole} = 1684.3 - 0.5 i \mev$.  The structure of the
resonance is reflected in $g_{j}$ and $w_{\rm pole}$, and recently
this is formulated in terms of the so-called
compositeness~\cite{Hyodo:2011qc, Hyodo:2013nka, Sekihara:2014kya,
  Sekihara:2015gvw}, which is defined as the norm of the two-body wave
function and measures the ``amount'' of the two-body component inside
the resonance.  In the present model, the $j$th channel compositeness
$X_{j}$ is expressed as
\begin{equation}
  X_{j} = - g_{j}^{2} \left [ \frac{d G_{j}}{d w} \right ] _{w = w_{\rm pole}} ,
  \quad
  Z \equiv 1 - \sum _{j} X_{j} 
  = - \sum _{j, k} g_{k} g_{j} 
  \left [ G_{j} \frac{d V_{j k}}{d w} G_{k} \right ] _{w = w_{\rm pole}} ,
  \label{eq:XZ}
\end{equation}
where we have also introduced the elementariness $Z$, which is given
by the rest of the component out of unity and measures contributions
from implicit channels that do not appear as explicit degrees of
freedom.  The last expression of the elementariness in our model is
obtained from the generalized Ward identity derived in
Ref.~\cite{Sekihara:2010uz}.  In Table~\ref{tab:1} we show the values
of the compositeness and elementariness as well as the coupling
constants for the $\Xi (1690)^{0}$ resonance in our model.  Since the
sum of the $K^{-} \Sigma ^{+}$ and $\bar{K}^{0} \Sigma ^{0}$
compositeness is almost unity, we can see that $\Xi (1690)^{0}$ is
indeed a $\bar{K} \Sigma$ molecular state.  Here we note that $\Xi
(1690)^{0}$ in our model has very small decay width $\sim 1 \mev$ with
a very small coupling constant to the $\pi \Xi$ channel.  This can be
understood by the structure of the Weinberg--Tomozawa term for $V_{j
  k}$.  Namely, in the Weinberg--Tomozawa term, the transition between
$\bar{K} \Sigma$ and $\eta \Xi$ is strong while the transitions
$\bar{K} \Sigma \leftrightarrow \bar{K} \Lambda$ and $\bar{K} \Sigma
\leftrightarrow \pi \Xi$ are forbidden and weak, respectively.  As a
consequence, the dynamically generated $\Xi (1690)$ in our model
cannot couple strongly to $\bar{K} \Lambda$ nor $\pi \Xi$.  This
solves the problem on the decay pattern of $\Xi (1690)$ with $J^{P} =
1/2^{-}$.

\begin{figure}[t]
  \centering
  \begin{minipage}{0.55\hsize}
    \Psfig{7.6cm}{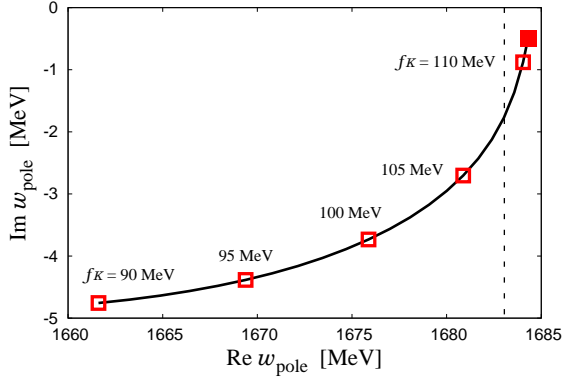}
  \end{minipage}
  \begin{minipage}{0.40\hsize}
    \caption{Shift of the $\Xi (1690)^{0}$ pole position $w_{\rm
        pole}$ with several values of the kaon decay constant $f_{K}$.
      The cases of $f_{K} = 90$, $95$, $100$, $105$, and $110 \mev$
      are plotted by the open boxes.  The case of the physical value
      $f_{K} = 1.2 f_{\pi} = 110.64 \mev$ is plotted by the filled
      box.  The vertical dotted line represents the $K^{-} \Sigma
      ^{+}$ threshold.}
    \label{fig:2}
  \end{minipage}
\end{figure}

Now let us compare the pole position of $\Xi (1690)$ in our model with
that obtained by the previous studies in the chiral unitary approach.
Namely, the $\Xi (1690)$ pole position was found at $1663 - 2 i
\mev$~\cite{GarciaRecio:2003ks} and at $1651 - 2 i
\mev$~\cite{Gamermann:2011mq}, both of which give binding energy of
several ten MeV from the $\bar{K} \Sigma$ threshold.  The difference
between our result and their ones can be understood by the $\bar{K}
\Sigma$ interaction strength.  First, while we use $f_{K} = 1.2
f_{\pi}$ with $f_{\pi} = 92.2 \mev$, to which we refer as the physical
kaon decay constant, the authors in Ref.~\cite{GarciaRecio:2003ks}
used $f_{K} = 90 \mev$, which makes the $\bar{K} \Sigma$ interaction
about $1.5$ times stronger than ours.  Actually, from
Fig.~\ref{fig:2}, in which we show the shift of the pole position by
changing the value of $f_{K}$, we can see that the case of $f_{K} = 90
\mev$ gives the binding energy $\approx 20 \mev$ due to a stronger
$\bar{K} \Sigma$ interaction.  We note that the physical $\Xi
(1690)^{0}$ pole (filled box) exists above the $K^{-} \Sigma ^{+}$
threshold but in the first Riemann sheet.  In this sense, strictly
speaking, the peak seen in Fig.~\ref{fig:1} is a cusp at the $K^{-}
\Sigma ^{+}$ threshold rather than a usual Breit--Wigner resonance
peak.  Second, in Ref.~\cite{Gamermann:2011mq} they introduced
channels with vector mesons, which would assist more the $\bar{K}
\Sigma$ interaction, and hence the mass of $\Xi (1690)$ could shift to
lower energies.

\begin{figure}[t]
  \centering
  \begin{minipage}{0.55\hsize}
    \Psfig{7.6cm}{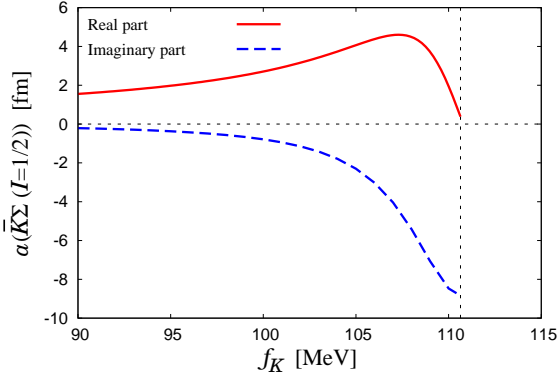}
  \end{minipage}
  \begin{minipage}{0.40\hsize}
    \caption{Scattering length of the $\bar{K} \Sigma ( I = 1/2 )$
      channel as a function of the kaon decay constant $f_{K}$
      obtained from the channels of charge zero.  The vertical dotted
      line represents the physical value of the kaon decay constant
      $f_{K} = 1.2 f_{\pi} = 110.64 \mev$.}
    \label{fig:3}
  \end{minipage}
\end{figure}

Finally, we discuss the spatial size of the $\Xi (1690)$ resonance,
which could be an important piece of the structure for
hadrons~\cite{Sekihara:2010uz, Sekihara:2008qk}.  Here we evaluate the
$\bar{K} \Sigma$ scattering length $a$, which is related to the mean
squared distance between two constituents $\langle r^{2} \rangle$.
Actually, in quantum mechanics we have $a^{2} \approx 2 \langle r^{2}
\rangle$ if the state is weak binding and is dominated by the two-body
component, i.e., $X \approx 1$~\cite{Weinberg:1965zz}.  The scattering
length in channel $j$, $a ( j )$, is calculated as $a ( j ) \equiv
M_{j} T_{j j} ( M_{j}^{\rm th} ) / (4 \pi M_{j}^{\rm th})$ with
$M_{j}$ and $M_{j}^{\rm th}$ being the baryon mass and threshold in
channel $j$, respectively.  In Fig.~\ref{fig:3} we show the scattering
length $a ( \bar{K} \Sigma ( I = 1/2 ) )$, which is obtained as $2 a (
K^{-} \Sigma ^{+} ) - a ( \bar{K}^{0} \Sigma ^{0} )$, by changing the
value of the kaon decay constant $f_{K}$.  From the figure, we can see
that the real part of the scattering length $a ( \bar{K} \Sigma ( I =
1/2 ) )$ becomes large as $f_{K}$ increases and the pole approaches
the $K^{-} \Sigma ^{+}$ threshold (see also Fig.~\ref{fig:2}).  In
particular, around $f_{K} = 105 \mev$ the scattering length becomes
$\approx 4 \fm$, which indicates that $\Xi (1690)$ would be a diffuse
system composed of $\bar{K}$ and $\Sigma$.  At the physical value of
the kaon decay constant, however, $\Xi (1690)$ exists above the $K^{-}
\Sigma ^{+}$ threshold, and the real part of the scattering length
becomes much small while its imaginary part is negatively large.

\section{Conclusion}

We have shown that the $\Xi (1690)$ resonance can be dynamically
generated in the $s$-wave $\bar{K} \Sigma$-$\bar{K} \Lambda$-$\pi
\Xi$-$\eta \Xi$ coupled-channels chiral unitary approach.  The $\Xi
(1690)$ resonance appears near the $\bar{K} \Sigma$ threshold.  In our
model, we can describe the $\Xi (1690)$ resonance as a $\bar{K}
\Sigma$ molecular state and reproduce well the experimental data,
including the decay pattern of the $\Xi (1690)$.  

We think that, in addition to the attractive interaction between kaon
and hadrons in chiral dynamics, the relatively heavy kaon mass
compared to the pion mass would be essential to the appearance of
hadronic molecules in the spectrum of multi-strangeness baryons such
as $\Xi (1690)$ in the present study, $\bar{K} \bar{K} N$ predicted in
Refs.~\cite{KanadaEn'yo:2008wm, Shevchenko:2015oea}, and so on.

This work is partly supported by the Grants-in-Aid for Scientific
Research from MEXT and JSPS (No.~15K17649, 
No.~15J06538
).

\end{document}